\documentclass[12pt,a4paper]{article}
\pdfoutput=1

\usepackage{amsmath}
\usepackage{amssymb}
\usepackage{epsfig}
\usepackage{graphicx}
\usepackage[numbers,sort&compress]{natbib}
\usepackage{color}
\usepackage[colorlinks=true
,urlcolor=blue
,anchorcolor=blue
,citecolor=blue
,filecolor=blue
,linkcolor=blue
,menucolor=blue
,pagecolor=blue
,linktocpage=true
,pdfproducer=medialab
,pdfa=true
]{hyperref}

\numberwithin{equation}{section}

\def\CA{{\cal A}}

\def\CL{{\cal L}}

\def\CN{{\cal N}}
\def\CO{{\cal O}}

\newcommand{\gsim}{ \mathop{}_{\textstyle \sim}^{\textstyle >} }
\newcommand{\lsim}{ \mathop{}_{\textstyle \sim}^{\textstyle <} }
\newcommand{\vev}[1]{ \left\langle {#1} \right\rangle }

\def\SO{\mathop{\rm SO}}

\def\SU{\mathop{\rm SU}}
\def\U{\mathrm{U}}
\def\Sp{\mathop{\rm Sp}}

\def\beq#1\eeq{\begin{align}#1\end{align}}
\begin{document}
\baselineskip 0.7cm
\begin{titlepage}

\begin{flushright}
%****
\end{flushright}

\vskip 1.35cm
\begin{center}
{\Large \bf 
Notes on natural inflation}
\vskip 1.2cm

Kazuya Yonekura %${}^{1}$
\vskip 0.4cm

{\it
%$^1$ 
School of Natural Sciences, Institute for Advanced Study, Princeton, NJ 08540, USA.
}

\vskip 1.5cm

\renewcommand{\baselinestretch}{1.1}  
\abstract{
In the so-called natural inflation, an axion-like inflaton is assumed to have a cosine-type periodic potential.
This is not the case in a very simple model in which the axion-like inflaton is coupled to an $\SU(N)$ (or other) pure Yang-Mills,
at least in the large $N$ limit as pointed out by Witten. It has a multi-valued potential, which is effectively quadratic, i.e., there is only a mass
term in the large $N$ limit. Thanks to this property, chaotic inflation can be realized more naturally with the decay constant
of the axion-like inflaton less than the Planck scale.
We demonstrate these points explicitly by using softly broken ${\mathcal N}=1$ Super-Yang-Mills which allows us to treat finite $N$.
This analysis also suggests that moderately large gauge groups such as $E_{8}$ are good enough with a Planck scale decay constant.
}
\end{center}
\end{titlepage}
\setcounter{page}{1}

%%%%%%%%%%%%%%%%%%%%%%%%%%%%%%%%%%%%%%%%%%%%%%%%%%%%%%%%%%%%%%%%%%%%%%%%
%\hrule\tableofcontents\bigskip\medskip\hrule
%%%%%%%%%%%%%%%%%%%%%%%%%%%%%%%%%%%%%%%%%%%%%%%%%%%%%%%%%%%%%%%%%%%%%%%%
\renewcommand{\baselinestretch}{1.3} \normalsize
%%%%%%%%%%%%%%%%%%%%%%%%%%%%%%%%%%%%%%%%%%%%%%%%%%%%%%%%%%%%%%%%%%%%%%%%
\section{Introduction}
%%%%%%%%%%%%%%%%%%%%%%%%%%%%%%%%%%%%%%%%%%%%%%%%%%%%%%%%%%%%%%%%%%%%%%%%
The recent announcement of the discovery of large tensor-to-scalar ratio $r$ of order $ \CO(0.1)$ by BICEP2 \cite{Ade:2014xna} suggests that 
the inflation scale is very high.\footnote{See e.g., \cite{Baumann:2009ds} for a review of inflation.} 
One of the scenarios which realizes such a large $r$
is the simplest chaotic inflation \cite{Linde:1983gd} with a Lagrangian
\beq
\CL=\frac{M_{\rm pl}^2}{2}R-\frac{1}{2} (\nabla \phi)^2-\frac{1}{2}m^2 \phi^2. \label{eq:originalchaos}
\eeq
where $M_{\rm pl} \simeq 2.4 \times 10^{18}~{\rm GeV}$ is the reduced Planck scale, $R$ is the Ricci scalar and $\phi$ is the inflaton.
Because the inflaton vacuum expectation value (VEV) must start from a very large value $\phi \gsim 15   M_{\rm pl}$
to realize the e-folds larger than $60$, the form of \eqref{eq:originalchaos} must be obeyed to a very good accuracy,
e.g., we cannot have corrections of order $\CO(1) \times \phi^n/M_{\rm pl}^{n-4}$.
Assuming that this chaotic inflation scenario is really the case, it is a theoretical challenge to find a reason why \eqref{eq:originalchaos}
is obeyed to such a good accuracy. 

It is pointed out in \cite{Freese:1990rb} that if the inflaton is an axion-like field $a$, then the shift symmetry $a \to a+{\rm const.}$
protects the inflaton potential from quantum corrections and hence the simple form of the potential becomes natural.
Such an idea has been realized in string theory \cite{Silverstein:2008sg,McAllister:2008hb} using axion monodromy.
However, string compactifications are far from unique, and little is known about the case in which supersymmetry is broken at the compactification
scale (e.g., internal manifold is not a Calabi-Yau). So it is also worth studying more bottom up approach using phenomenological Lagrangians.

The shift symmetry of the axion $a \to a+{\rm const.}$ is assumed to be preserved only perturbatively, but the symmetry $a \to a +2 \pi$ is exact.
Then, one might think that the axion potential is of the form
\beq
V(a)=\sum_{k=1}^\infty \left(v_k \cos( ka )+u_k \sin (ka) \right).\label{eq:cosinepot}
\eeq
If this is the case, the axion decay constant $f $ must be very large, $f >\CO(10) \times M_{\rm pl}$ to realize
the large tensor-to scalar ratio for the inflaton $\phi=f a$ (see e.g., \cite{Freese:2014nla,Creminelli:2014oaa} for recent discussions), which might be difficult
to realize in string theory~\cite{Banks:2003sx}.

What is the simplest scenario in which the axion potential is generated?  
The potential is assumed to be generated by nonperturbative effects. Within the low energy field theory, probably the simplest
theory is to use a pure Yang-Mills of some simple gauge group $G$ with the Lagrangian (from now on we omit Einstein-Hilbert term)
\beq
\CL=-\frac{1}{2}f^2 (\partial a)^2 -\frac{1}{4g^2} F_{\mu\nu}F^{\mu\nu}+ \frac{a}{32\pi^2}F_{\mu\nu}\tilde{F}^{\mu\nu},
\eeq
where $f$ is the decay constant of the axion, $g$ is the gauge coupling, $F_{\mu\nu}$ is the gauge field strength and 
$\tilde{F}^{\mu\nu}=\frac{1}{2}F_{\rho\sigma} \epsilon^{\rho\sigma\mu\nu}$. 
(Sums over adjoint index of the field strength are implicit.)
This theory might be natural because gauge fields and axions are what we expect to be perturbatively massless even without supersymmetry in string
compactifications.

Now, the important point is that the pure Yang-Mills is a strongly coupled theory where weak coupling instanton computations break down. 
Then, the potential \eqref{eq:cosinepot} is not true!
At least in the large $N$ limit of $G=\SU(N)$ theory, the potential is multi-valued due to the existence of many metastable vacua
as discussed by Witten~\cite{Witten:1979vv,Witten:1980sp,Witten:1998uka}. A continuous and smooth piece of the potential is of the form
\beq
V(a) \to \frac{1}{2} \Lambda^4 a^2~~~(N \to \infty),
\eeq
where $\Lambda$ is the dynamical scale of the gauge theory.
Therefore, this is a very simple field theory realization~\cite{Kaloper:2011jz,Dubovsky:2011tu}~\footnote{
See also e.g., \cite{Kaloper:2008fb,Berg:2009tg,Kaloper:2014zba,Harigaya:2014eta} for similar models of monodromy inflation.
There are also different mechanisms which use large $N$ for natural inflation, such as 
N-flation~\cite{Dimopoulos:2005ac} and M-flation~\cite{Ashoorioon:2009wa}.} 
of axion monodromy inflation 
scenario~\cite{Silverstein:2008sg,McAllister:2008hb}. In particular, the decay constant need not be super-Planckian.
In the pure Yang-Mills, it is not clear how large $N$ is required to realize the chaotic inflation unless we solve the 
finite $N$ strong dynamics of this theory. In this note we study a slightly modified theory, which is an $\CN=1$ supersymmetric
Yang-Mills (SYM) with the supersymmetry softly broken by the gaugino mass.\footnote{The idea of using strong dynamics for inflation
has also appeared in e.g., \cite{Harigaya:2012pg,Harigaya:2014sua}.} 
This theory allows us a reliable calculation
even for finite $N$, and may have its own interest if we try to embed the theory to low (or intermediate) scale supersymmetry scenario
and string theory.

The organization of the rest of this note is as follows. In section~\ref{sec:2}, 
we briefly review the large $N$ dynamics~\cite{Witten:1979vv,Witten:1980sp,Witten:1998uka}. In section~\ref{sec:3},
we discuss the softly broken $\CN=1$ SYM. Then conclusion and discussions are given in section~\ref{sec:4}.

\section{Large $N$ dynamics and the $\theta$ angle}   \label{sec:2}
Here we review the dependence on the $\theta$ angle in large $N$ gauge theory.
See \cite{Witten:1979vv,Witten:1980sp,Witten:1998uka} for more detailed and convincing explanations.

Let us consider a pure $G=\SU(N)$ Yang-Mills theory with the Lagrangian
\beq
\CL=-\frac{1}{4g^2} F_{\mu\nu}F^{\mu\nu}+ \frac{\theta}{32\pi^2}F_{\mu\nu}\tilde{F}^{\mu\nu}, \label{eq:pureYM}
\eeq
where $\theta$ is the usual theta angle. We would like to consider the vacuum energy $V(\theta)$ as a function of $\theta$.

We consider the usual large $N$ expansion 
where the 't~Hooft coupling $\lambda=Ng^2$ is kept fixed and $N$ is taken to be very large.
The vacuum energy is of order $N^2$, simply because there are order $N^2$ gluons whose loops contribute to the vacuum energy.
Then, in the large $N$ expansion, one can see that $\vev{ ( \frac{N}{\lambda}F_{\mu\nu}F^{\mu\nu})^n} =\CO(N^2)~(n=1,2,\cdots)$.
Since $F_{\mu\nu}\tilde{F}^{\mu\nu}$ has the same color structure as $F_{\mu\nu}F^{\mu\nu}$, we get 
\beq
\vev{(NF_{\mu\nu}\tilde{F}^{\mu\nu})(x_1) \cdots  (NF_{\mu\nu}\tilde{F}^{\mu\nu})(x_n)} =\CO(N^2).\label{eq:topologicalcorr}
\eeq
The vacuum energy $V(\theta)$ is computed by Euclidean path integral as
\beq
\exp\left[ -V(\theta) {\rm Vol}({\mathbb R}^4) \right]=\int [DA_{\mu}] \exp(-S_{\rm E}), \label{eq:eucpath}
\eeq
where $S_E$ is the Euclidean action obtained from Wick rotation of \eqref{eq:pureYM}, and ${\rm Vol}({\mathbb R}^4)$ is the volume of space-time
introduced to make the result finite.
Then we can expand the exponential $ \exp(-S_{\rm E})$ in terms of $\theta$ and compute the vacuum energy.
Using \eqref{eq:topologicalcorr}, we can see that the vacuum energy is of the form
\beq
V(\theta)=N^2 h(\theta/N) +\cdots, \label{eq:generalpot}
\eeq
where $h$ is an order one function which is independent of $N$.

Now let's take the limit $N \to \infty$. In this limit, the $\theta/N$ becomes very small (if we keep $\theta$ fixed) and we can expand the function $h$.
The CP symmetry (assuming it is not spontaneously broken in the vacuum under consideration) tells us that $h(\theta/N)=h(-\theta/N)$, so we get 
\beq
V(\theta) \to \frac{1}{2}\Lambda^4 \theta^2~~~(N \to \infty), \label{eq:largeNtheta}
\eeq
where $\Lambda$ is the dynamical scale obtained from the second derivative of the function $h$.
This second derivative is positive since the Euclidean path integral \eqref{eq:eucpath} is maximized when $\theta$ is zero
due to the inequality $\int [DA] e^{-S_E} \leq \int [DA] |e^{-S_E}|$ ~\cite{Vafa:1984xg}.

How can the above potential \eqref{eq:largeNtheta} be consistent with the periodicity under $\theta \to \theta +2\pi$?
The solution to this problem is that there are many metastable vacua in the theory, parametrized by an integer $k$.
In the vacuum specified by $k$, the potential is given by
\beq
V_k(\theta)=N^2 h((\theta+2\pi k)/N) \to \frac{1}{2}\Lambda^4 (\theta +2\pi k)^2.\label{eq:metavacua}
\eeq
The true vacuum is determined by $\min_{k} V_k(\theta)$. However, other vacua are also long lived in the large $N$ limit 
(see section~\ref{sec:3}). These vacua can be used for chaotic inflation by replacing $\theta \to a$ as discussed in the introduction.

The only subtlety in the above discussion is that $F_{\mu\nu}\tilde{F}^{\mu\nu}$ is a total derivative, and one might think that 
the integral $\int d^4 x F_{\mu\nu}\tilde{F}^{\mu\nu}$ is zero in the absence of instantons.
However, this is not the case in strongly coupled theories due to strong infrared divergences.\footnote{A possible mechanism in which
infrared divergences play a role is as follows. The $F_{\mu\nu}\tilde{F}^{\mu\nu}$ can be written as 
$F_{\mu\nu}\tilde{F}^{\mu\nu}=\partial_\mu K^\mu$. 
The integral $\int d^4 x F_{\mu\nu}\tilde{F}^{\mu\nu}$ takes the zero momentum mode, so naively, in momentum space we get
$F_{\mu\nu}\tilde{F}^{\mu\nu}=ip_\mu K^\mu \to 0~(p_\mu \to 0)$. However, suppose that correlators of $K^\mu$ have poles in the infrared, e.g.,
$\vev{K^\mu K^\nu} \sim \eta^{\mu\nu} \frac{1}{p^2}$. Then we get $\vev{p_\mu K^\mu p_\nu K^\nu} \sim 1$, which is nonzero for $p_\mu \to 0$.
One can explicitly check that this mechanism happens in two dimensional free Maxwell theory
$S=\int d^2x [(-1/4e^2)F_{\mu\nu}F^{\mu\nu}+(\theta/4\pi) \epsilon^{\mu\nu} F_{\mu\nu}]$. Computing the vacuum energy
using the formula \eqref{eq:eucpath} in the way described above,
one gets $V(\theta)=\frac{e^2}{2} (\frac{\theta}{2\pi})^2$, which agrees with \cite{Coleman:1976uz}.
}
Although direct argument is difficult in four dimensions, one can ask a similar question in two-dimensional sigma models
where it is explicitly demonstrated that the $\theta$ dependence comes from a resummation of Feynman diagrams~\cite{Witten:1978bc, D'Adda:1978uc}.
In fact, the theta dependence very similar to that of \eqref{eq:largeNtheta}, \eqref{eq:metavacua} was first observed 
in massive Schwinger model in \cite{Coleman:1976uz}.
See \cite{Witten:1979vv,Witten:1980sp} for how these arguments are consistent with the solution of the $\U(1)$ problem in QCD.
(The fact that the $\U(1)$ problem is solved not by instantons as discussed by 't~Hooft~\cite{'tHooft:1976up}, but by the way 
suggested in \cite{Witten:1979vv,Witten:1980sp} is also confirmed by a holographic QCD model \cite{Sakai:2004cn}.)

The above arguments are rather abstract, but they are explicitly demonstrated in large $N$ QCD with small quark masses using chiral 
Lagrangian~\cite{Witten:1980sp},
and in a holographic model which is believed to have the same qualitative behavior as the pure Yang-Mills~\cite{Witten:1998uka}.
We discuss another example in the next section.

\section{Softly broken $\CN=1$ Super-Yang-Mills} \label{sec:3}
We consider an $\CN=1$ SYM of gauge group $\SU(N)$ with a small supersymmetry breaking gaugino mass.
The Lagrangian is
\beq
\CL=\frac{1}{g^2} \left(-\frac{1}{4} F_{\mu\nu}F^{\mu\nu}+i\bar{\lambda} \bar{\sigma}^\mu \partial_\mu \lambda \right)
+ \frac{\theta}{32\pi^2}F_{\mu\nu}\tilde{F}^{\mu\nu}+ \left(\mu N \frac{\lambda\lambda}{32\pi^2}+{\rm h.c.} \right),
\eeq
where $\lambda$ is the gaugino, and $\mu$ is a parameter which is related to the gaugino mass $m_\lambda$ at the tree level as
\beq
\mu =\frac{16 \pi^2}{Ng^2} m_\lambda.
\eeq
In the infrared where the theory is strongly coupled, $\mu \sim \CO(1) \times m_\lambda$ in the large $N$ counting.
This theory has the same large $N$ counting as the pure Yang-Mills, so the arguments of the previous section should apply to this theory.
We will see that this is indeed the case.

First, let us consider the case $\mu=0$ (see \cite{Intriligator:1995au} for a review). 
In this case, there is a ${\mathbb Z}_{2N}$ $R$-symmetry $\lambda \to e^{\pi i k/N}~(k=1,\cdots,2N)$.
In the IR, this discrete symmetry is spontaneously broken down to ${\mathbb Z}_2$ by the gaugino condensation,
\beq
\vev{N\frac{\lambda\lambda}{32\pi^2}}= \exp \left( i\frac{2 \pi  k +\theta}{N} \right) N^2\Lambda^3,
\eeq
where the integer $k=1,\cdots,N$ specifies the vacua, and $\Lambda$ is the dynamical scale normalized by this equation.
(This normalization of $\Lambda$ differs from that of \cite{Intriligator:1995au} so that it is consistent with the large $N$ counting.)
The presence of these $N$ vacua is required by the ${\mathbb Z}_{2N}$ symmetry and the gaugino condensation 
$\vev{\lambda\lambda} \neq 0$.\footnote{Although gaugino condensation is a strong coupling phenomenon, it can be argued very convincingly
by starting from weakly coupled supersymmetric QCD with $N_f=N-1$ flavors \cite{Affleck:1983mk}
and then decoupling quarks by giving masses (see \cite{Intriligator:1995au}).
It is also consistent with the Witten index~\cite{Witten:1982df,Witten:2000nv}.}
The $\theta$ dependence is required by the anomaly $\lambda \to e^{i\alpha} \lambda$, $\theta \to \theta +2N\alpha$ for $\alpha \in {\mathbb R}$.\footnote{
When we take $\theta \to a$, there is a continuous $\U(1)$ global symmetry under which the ``glueball'' $\lambda\lambda$ has charge 2 and
the exponential of the axion $e^{i a}$ has charge $2N$. This symmetry is explicitly broken by the gaugino mass. Then, the model discussed here
is a dynamical realization of the general mechanism discussed in \cite{Harigaya:2014eta,Harigaya:2014rga}. We thank M.~Ibe for pointing this out to us.}

Now let us turn on nonzero $\mu$. If $\mu$ is small enough, $\mu \ll \Lambda$, so that we can treat this parameter perturbatively,
the vacuum energy in the $k$-th vacuum is given by
\beq
V_k(\theta) &=-\vev{\mu N \frac{\lambda\lambda}{32\pi^2}+{\rm h.c.} }+{\rm const.} \nonumber \\
&=N^2 \mu \Lambda^3\left[ 1-\cos\left(  \frac{2 \pi  k +\theta}{N}\right) \right],\label{eq:gaugenopot}
\eeq
where we have tuned the cosmological constant. This potential is exactly of the form \eqref{eq:generalpot} and \eqref{eq:metavacua} 
predicted by the large $N$ argument.
Note also that there are many (or more precisely $N$) matastable vacua labelled by the integer $k$.
Therefore we have confirmed the predictions of the general arguments of large $N$ expansion.
Furthermore, the present argument have not used large $N$ expansion, so \eqref{eq:gaugenopot} is valid even for finite $N$ as long as $\mu$
is small enough.

For matastable vacua to be useful, we have to make sure that they are sufficiently long lived.
Let us estimate the vacuum decay rate of matastable vacua using thin-wall approximation~\cite{Coleman:1977py}\footnote{
We neglect gravitational corrections \cite{Coleman:1980aw} for simplicity. }, whose justification will be discussed later.
For concreteness we compute the decay rate from the $k=0$ vacuum to the $k=-1$ vacuum for $0<\theta <\pi N/2 $.

In the limit $\mu \to 0$, the two vacua are separated by a strongly coupled domain wall~\cite{Dvali:1996xe}.
The tension of the domain wall is believed to be BPS saturated and is given by~\cite{Dvali:1996xe}
\beq
T&=2N \left| \vev{\frac{\lambda\lambda}{32\pi^2}}_{k=0}- \vev{\frac{\lambda\lambda}{32\pi^2}}_{k=-1} \right| \nonumber \\
& \simeq 4 \pi N \Lambda^3. \label{eq:tension}
\eeq
This tension may not be significantly changed for nonzero $\mu$ as long as $\mu \ll \Lambda$.
Now the relevant ``classical'' configuration for the vacuum decay is the following.
The decay is produced by a nucleation of a bubble of radius $R$ in four dimensional Euclidean space. 
Inside $R$, there is a vacuum with $k=-1$, and outside it there is a vacuum with $k=0$, and they are separated by the domain wall. 
The classical action for this process is
\beq
S_E&=-\frac{\pi^2}{2} R^4 \left( V_{k=0}-V_{k=-1} \right) + 2 \pi^2 R^3 T \nonumber \\
& \simeq -\pi^3 N\mu \Lambda^3 R^4 \sin(\theta/N)+8\pi^3N \Lambda^3 R^3.
\eeq
Extremizing this action, we get 
\beq
S_E \simeq 432 \pi^3 N \frac{\Lambda^3}{ \mu^3 \sin^3 (\theta/N)},~~~~R \simeq \frac{6}{\mu \sin(\theta/N)}.
\eeq
The action has enhancement due to the factors $N$ and $\Lambda^3/\mu^3$.
The factor $N$ is expected to be a generic feature of large $N$ gauge theory; see \cite{Witten:1980sp,Witten:1998uka}.
A very crude estimation of the probability of decay within observable universe during inflation is given as
$P=[\Lambda^4 e^{-S_E}] \times [H^{-4} (e^{60})^3]$, where the first factor is the decay rate per volume,
and the second factor is the four dimensional volume ($H$ is the Hubble parameter during inflation) where it is taken into account that the e-folds of order 60 enhances the spatial volume
about $(e^{60})^3$. This probability $P$ can be less than one by taking $N$ and $\Lambda^3/\mu^3$ appropriately large.

In the above computation, we have used the thin-wall approximation. For this to be valid,
the ``thickness'' of the domain wall, which we denote as $\ell$, must be much smaller than the radius $R$.
In \cite{Witten:1997ep}, it is argued that domain walls in the large $N$ limit are kind of fundamental excitations rather than solitons 
of ``glueball effective action''. The reason is that the glueball effective action in the large $N$ limit has the general form
$\CL=N^2 f( G)$, where $G$ represents the glueball and $f$ is an order one function. Then, if the domain walls were solitons,
its tension would be of order $N^2$, but we can see from \eqref{eq:tension} that its tension is actually of order $N$.
So the domain walls are something similar to D-branes in closed string theory, and its thickness is negligibly small in the large $N$ limit.
Actually, in holographic duals of confining theories~\cite{Witten:1998zw,Witten:1998uka,Polchinski:2000uf,Klebanov:2000hb,Maldacena:2000yy},
domain walls are realized by D-branes. The ``thickness of a D-brane'' (measured very crudely by using the metric of black branes) is of
order $g_s \sim N^{-1}$ where $g_s$ is the string coupling. Therefore, we expect that the thickness is of order $\ell \sim (N\Lambda)^{-1}$.
Comparing with the $R$ obtained above, we get $\ell \ll R$, justifying the thin-wall approximation.

Now let us return to the axion inflation. We replace $\theta \to a$ and take the vacuum $k=0$ without loss of generality. 
Defining the inflaton as $\phi=f a$, its effective action is now given by
\beq
\CL=-\frac{1}{2} (\partial \phi)^2-N^2 \mu \Lambda^3\left[ 1-\cos\left(  \frac{\phi}{N f}\right) \right].\label{eq:infeffective}
\eeq
This is almost the same as the original natural inflation action of \cite{Freese:1990rb}.
However, now the decay constant is effectively enhanced to $Nf$.
Now the constraint is that $Nf \gsim \CO(10) \times M_{\rm pl}$, and hence the decay constant can be less that the Planck scale 
if $N \gsim \CO(10)$.

Let us comment on the case of more general gauge group $G$. In this case, the gaugino condensation is given by
\beq
\vev{\frac{\lambda\lambda}{32\pi^2}} \propto \exp \left( i\frac{2 \pi  k +\theta}{h} \right) ,
\eeq
where $h$ is the dual Coxeter number of the gauge group $G$. Correspondingly, the inflaton action is given by 
\eqref{eq:infeffective} with the replacement $N \to h$.
For example, the dual Coxeter number of the gauge groups  $\SO(N)~(N \geq 5)$, $\Sp(N)$, $E_6$, $E_7$ and $E_8$ are given by
\beq
h(\SO(N))=N-2,~~~h(\Sp(N))=N+1,~~~h(E_6)=12,~~~h(E_7)=18,~~~h(E_8)=30. 
\eeq
Therefore, the gauge groups such as $E_{8}$ can be consistent with $f \lsim M_{\rm pl}$.
This is valid as long as $\mu \ll \Lambda $.  
We are not sure what happens when $\mu \to \infty$, i.e., pure bosonic Yang-Mills, but it is interesting to speculate that
the above scenario is qualitatively true even in bosonic Yang-Mills.

Before closing this section, we would like to comment on the so-called species problem~\cite{ArkaniHamed:2005yv,Distler:2005hi,Dimopoulos:2005ac}.
In general, the cutoff scale $\Lambda_{\rm UV}$ of the theory is not necessarily the same as the Planck scale $M_{\rm pl}$.
In particular, if the degrees of freedom of the theory is of order $N^2$, there can be quantum corrections to the Planck scale such that
\beq
M_{\rm pl}^2 \sim M_{\rm pl,0}^2 \pm N^2\frac{\Lambda_{\rm UV}^2}{16 \pi^2},
\eeq
where $M_{\rm pl,0}$ is the ``bare" Planck scale. Therefore, if $N$ is very large, the cutoff scale is much smaller than 
the Planck scale as $\Lambda_{\rm UV} \sim 4\pi M_{\rm pl} / N$. If the decay constant $f$ is assumed to be smaller than
the scale $\Lambda_{\rm UV}/4\pi$, we get $Nf \lsim N\Lambda_{\rm UV}/4\pi \sim M_{\rm pl} $ and we could not satisfy
the condition $Nf \gsim \CO(10) M_{\rm pl}$ required by inflation.

In our case, we need $N$ to be not infinity, but to be of order $\CO(10)$. Thus it is a numerical question whether or not 
our scenario is possible, and that should be addressed in a UV complete theory of gravity such as string theory.
However, we would like to make a simple observation from the low energy effective field theory point of view.
As an example, suppose that the kinetic term of the gaugino is modified from 
$g^{-2}i\bar{\lambda} \bar{\sigma}^\mu \partial_\mu  \lambda$
to $g^{-2}i\bar{\lambda} \bar{\sigma}^\mu (\partial_\mu \lambda+\partial_\mu a \lambda)$ so that the axion has a derivative coupling to the gaugino.
Then, gaugino loops may generate quantum corrections to the decay constant as
\beq
f^2 \sim f^2_0 \pm N^2\frac{\Lambda_{\rm UV}^2}{16 \pi^2}.
\eeq
Therefore, even if the bare decay constant $f_0$ is less than the cutoff scale $\Lambda_{\rm UV}$, 
we may still get $f \sim N \Lambda_{\rm UV} /4\pi \sim M_{\rm pl}$ in the low energy effective theory.
To the best of the author's knowledge, there is no concrete low energy argument which forbids a decay constant in the range 
$(\Lambda_{\rm UV}/4\pi) \lsim f  \lsim M_{\rm pl}$. 
For example, the weak gravity conjecture of \cite{ArkaniHamed:2006dz} seems to require only $f \lsim M_{\rm pl}$.
Note that effective field theory is valid even if $f \gsim \Lambda_{\rm UV}/4\pi $, and as long as we get $f$ near $M_{\rm pl}$, our scenario works.

\section{Conclusion and discussions} \label{sec:4}
We have revisited the potential of the axion-like inflaton $a$ in natural inflation scenario.
Because of the periodicity $a \cong a +2\pi$, it is usually assumed to be
\beq
\CL=-\frac{1}{2}f^2 (\partial a)^2 -\Lambda^4[1-\cos (a)].
\eeq
However, if the axion potential is generated by nonperturbative dynamics of strongly coupled theory such as pure Yang-Mills, 
this is not the case. There are many metastable vacua, and the potential in each vacuum is not periodic under $a \to a +2\pi$ similar to axion monodromy 
inflation.
We have demonstrated this fact explicitly by using softly broken $\CN=1$ Super-Yang-Mills.
For example, if the gauge group is $\SU(N)$, the low energy Lagrangian of the inflaton is now
\beq
\CL=-\frac{1}{2}f^2 (\partial a)^2 - N^2 \mu \Lambda^3 \left[1-\cos\left(\frac{a+2\pi k}{N} \right)\right].
\eeq
where $\mu$ is the small gaugino mass and $k=0,1,\cdots,N-1$ is an integer specifying the metastable vacua.
Inflation can happen in any of the vacua, say $k=0$, and the decay constant $f$ can be less than the Planck scale if $N$ is moderately large.
Although we have concentrated on the softly broken Super-Yang-Mills, it is a rather generic feature of strongly coupled gauge theories
in which the vacuum energy depends on the theta angle.

In our scenario, supersymmetry is not essential at all; pure bosonic Yang-Mills is of course nonsupersymmetric, and softly broken Super-Yang-Mills
is just a gauge theory with a massive majorana fermion in the adjoint representation (i.e., there is no need to fine-tune coupling constants).
However, it is also interesting to embed the theory to supergravity and string theory. 
Let us briefly discuss these directions, leaving the details for future work. 

To embed the scenario to supergravity, we minimally need $\CN=1$ SYM coupled to an axion superfield $\CA=s+ia+\cdots$, where $s$ is the ``saxion''.
Even in supergravity, the shift symmetry $\CA \to \CA+ic~(c \in {\mathbb R})$ protects the axion potential from several corrections~\cite{Kawasaki:2000yn}.
Then, what is necessary is to generate the large gaugino mass and stabilize the saxion $s$. The simplest scenario may be that 
the saxion is stabilized by gravity mediation and the gaugino mass is generated by anomaly mediation.
The anomaly mediation generates the gaugino mass as \cite{Giudice:1998xp,Randall:1998uk} $\mu=3m_{3/2}$, where $m_{3/2}$ is the gravitino mass.
Because the inflaton mass $m_a \simeq \sqrt{\mu \Lambda^3}/f$ must be of order $10^{13}~{\rm GeV}$ to generate the observed density fluctuations,
we need a rather high scale supersymmetry breaking. 
(See, e.g., \cite{Hall:2013eko,Hall:2014vga} for recent discussions on such intermediate scale supersymmetry.
See also \cite{Harigaya:2014ola} where an interesting model is proposed which makes low scale supersymmetry compatible with 
our scenario.)
However, if the gaugino mass and the saxion mass are generated by other mechanisms or if the origin of their masses during inflation are 
different from that of today, the supersymmetry breaking scale may be lowered. 
\\
\\
{\bf Note added:}\\
While this work was being completed, the paper \cite{Dine:2014hwa} appeared on arXiv, which has some overlap with this note.

%%%%%%%%%%%%%%%%%%%%%%%%%%%%%%%%%%%%  acknowledgements
\subsection*{Acknowledgments}
%\paragraph{Acknowledgement.}---
%%%%%%%%%%%%%%%%%%%%%%%%%%%%%%%%%%%%  
The author would like to thank M.~Yamazaki for helpful discussions. 
The work of K.Y. is supported in part by NSF Grant PHY-0969448.

%\appendix
%%%%%%%%%%%%%%%%%%%%%%%%%%%%%%%%%%%%%%%%%%%%%%%%%%%%%%%%%%%%%%%%%%%%%%%%%%%%%
%\section{aaa}
%\label{appsec:}

\bibliographystyle{JHEP}

\renewcommand{\baselinestretch}{1.1}  
\let\oldthebibliography=\thebibliography
\let\endoldthebibliography=\endthebibliography
\renewenvironment{thebibliography}[1]{%
\begin{oldthebibliography}{#1}
\small%
\raggedright%
\setlength{\itemsep}{5pt plus 0.2ex minus 0.05ex}%
}%
{%
\end{oldthebibliography}%
}

\bibliography{ref}

\end{document}